\documentclass[10pt,twocolumn,a4paper,superscriptaddress,aps,pra,longbibliography]{revtex4-1}

\usepackage{amsmath,amssymb,amsfonts}

\usepackage{graphicx}
\usepackage[colorlinks,
	linkcolor=red,
	citecolor=red,
	urlcolor=red]{hyperref}

\usepackage{cleveref}

\def\bra#1{\mathinner{\langle{#1}|}}
\def\ket#1{\mathinner{|{#1}\rangle}}

\newcommand{\pw}[1]{ \cdot 10^{#1}}

\renewcommand{\t}[1]{\mathrm{#1}}
\newcommand{\op}[1]{\hat{{#1}}}
\newcommand{\mean}[1]{\left\langle {#1} \right\rangle}

\newcommand{\nocontentsline}[3]{}
\newcommand{\tocless}[2]{\bgroup\let\addcontentsline=\nocontentsline#1{#2}\egroup}

\begin{document}

\title{Preparation and decay of a single quantum of vibration at ambient conditions}

\author{Santiago Tarrago Velez}
\author{Kilian Seibold}
\author{Nils Kipfer}
\author{Mitchell D. Anderson}
\affiliation{Institute of Physics, Ecole Polytechnique F\'{e}d\'{e}rale de Lausanne (EPFL), CH-1015 Lausanne, Switzerland}
\author{Vivishek Sudhir}
\affiliation{LIGO Laboratory, Massachusetts Institute of Technology, Cambridge, MA 02139, USA}
\affiliation{Department of Mechanical Engineering, Massachusetts Institute of Technology, Cambridge, MA 02139, USA}
\author{Christophe Galland}
\affiliation{Institute of Physics, Ecole Polytechnique F\'{e}d\'{e}rale de Lausanne (EPFL), CH-1015 Lausanne, Switzerland}

\date{\today}


\begin{abstract}

A single quantum of excitation of a mechanical oscillator is a textbook example of the principles of quantum physics. 
But mechanical oscillators, despite their pervasive presence in nature and modern technology, 
do not generically exist in an excited Fock state. 
In the past few years, careful isolation of GHz-frequency nano-scale oscillators has allowed experimenters to prepare 
such states at milli-Kelvin temperatures. 
These developments illustrate the tension between the basic predictions of quantum mechanics -- which should apply to all mechanical 
oscillators even at ambient conditions -- and the extreme conditions required to observe those predictions.
We resolve the tension by creating a single Fock state of a 40~THz vibrational mode in a crystal at room temperature 
and atmospheric pressure. 
After exciting a bulk diamond with a femtosecond laser pulse and detecting a Stokes-shifted photon, the Raman-active vibrational mode is prepared in the Fock state $\ket{1}$ with $98.5\%$ probability.
The vibrational state is then mapped onto the anti-Stokes sideband of a subsequent pulse, which when subjected to a Hanbury-Brown-Twiss intensity correlation measurement reveals the sub-Poisson number statistics of the vibrational mode. 
By controlling the delay between the two pulses we are able to witness the decay of the vibrational Fock state over 
its $3.9$ ps lifetime at ambient conditions.
Our technique is agnostic to specific selection rules, and should thus be applicable to any Raman-active medium, opening
a new general approach to the experimental study of quantum effects related to vibrational degrees of freedom in molecules and solid-state systems.

\end{abstract}

\maketitle

\tocless\section{Introduction}

\begin{figure*}[t!]
\centering
\includegraphics[width=\textwidth]{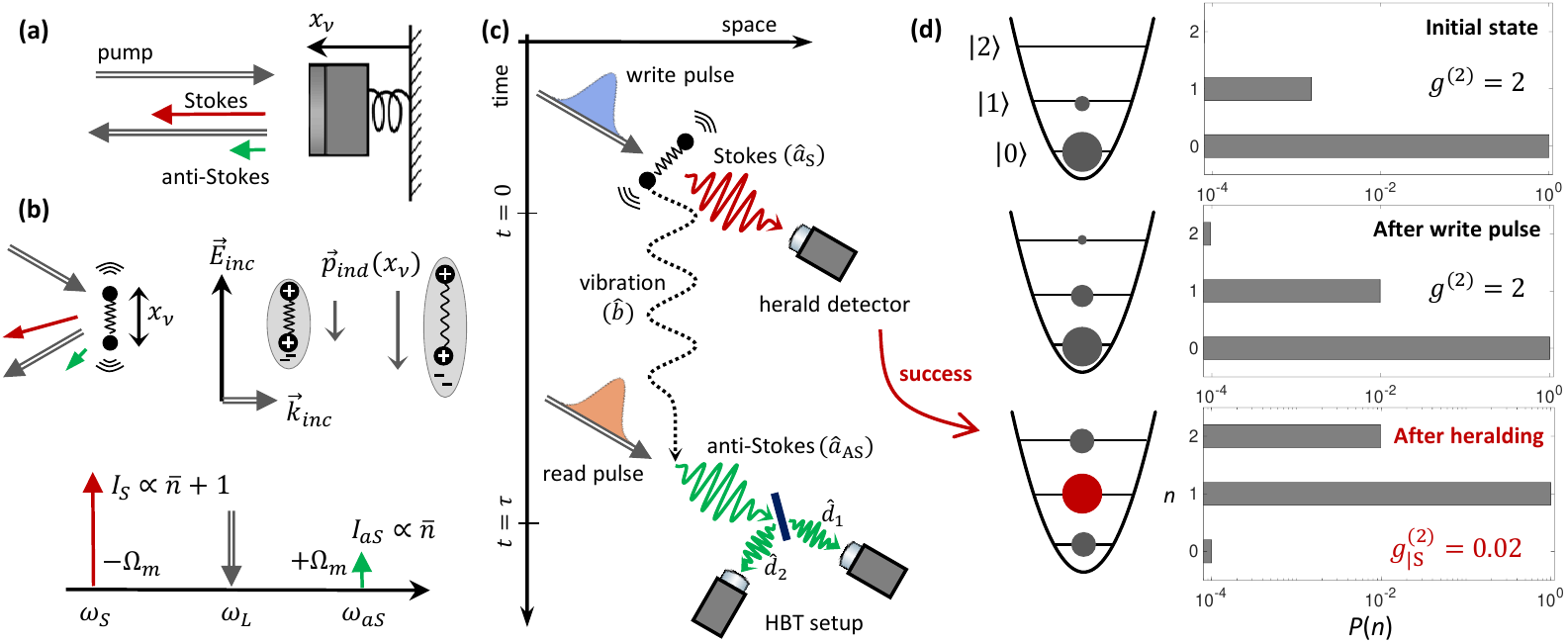}
\caption{\label{fig:intro} \textbf{Concept of the experiment.}
(a) When monochromatic light reflects from an oscillating mirror, it acquires two Raman sidebands due to phase modulation (to first order in interaction strength). 
(b) When light interacts with a polarizable oscillator, whose induced polarization $\vec{p}_\t{ind}$ depends on its internal coordinate $x_\nu$, it undergoes a phase modulation at the oscillator frequency, leading to the appearance of Stokes and anti-Stokes Raman sidebands in the scattered light. The anti-Stokes/Stokes intensity ratio is proportional to $\frac{\bar{n}}{\bar{n}+1}$ where $\bar{n}$ is the mean excitation number (or occupancy) of the vibrational mode. (c) Time evolution of a single repetition of the experiment showing the interaction of the write pulse, the subsequent detection of Stokes photons for heralding, and the final read pulse followed by measurement of two-photon correlation in the emitted anti-Stokes light. 
(d) Evolution of the probability $P(n)$ of finding the vibrational mode in the $n^{\text{th}}$ energy eigenstate during the different steps. The system is initially in a thermal state with $\bar{n}= 1.5 \pw{-3}$ (for a mode frequency of 40~THz at 295~K). After the write pulse, the marginal state of the vibration is also thermal with 
$\bar{n}= \frac{p}{1-p} = 10^{-2}$ in this example (here $p$ is the interaction probability). 
Finally, after heralding, the distribution becomes peaked at $n=1$. The residual vacuum component is due to detection noise. For each distribution, we give the corresponding value of the vibrational mode's intensity correlation function $g^{(2)}$. 
}
\end{figure*}

The observation of inelastic scattering of photons from ensembles of atomic-scale particles was an early triumph of quantum theory. Within a few years, experiments by Compton \cite{compton1923} and Raman \cite{raman1928} showed
that photons can exchange energy and momentum with material particles in the manner described by quantum mechanics. At optical frequencies, Raman scattering, the dominant effect, is an expression of the universal idea that a mechanical vibration phase-modulates the outgoing light, resulting in two scattered sidebands (Fig.~\ref{fig:intro}a,b). In a quantum description, the upper (``anti-Stokes'') sideband arises from the annihilation of a quantum of vibration, while the lower (``Stokes'') arises from its creation (Fig.~\ref{fig:intro}c).

Leveraging the Raman interaction, a variety of pump-probe measurements have been implemented to study 
vibrational dynamics in crystals and molecules. For example, incoherent phonons generated by the decay of electron-hole pairs can  be probed by time-resolved anti-Stokes scattering \cite{vonderlinde1980,kash1985,song2008,jiang2018}. Several techniques have also been developed to study coherent states of the vibrational modes. The most popular is time-resolved coherent anti-Stokes Raman scattering, where a large coherent phonon population is excited by a pair of laser pulses, 
and its classical decay is probed by a delayed pulse \cite{mukamel1995}. Another technique -- transient coherent ultrafast phonon spectroscopy -- uses the interference of the Stokes photons from the spontaneous Raman scattering of two coherent pumps to determine the decoherence of the vibrational mode \cite{waldermann2008,Lee2010}.
While these techniques reveal the time scales over which the vibration decays or loses its phase coherence,
observing single quanta of the vibration itself has proved far more elusive. 

To illustrate the difficulty, consider that
on the one hand, internal vibrational modes of crystals and molecules with oscillation frequencies in the 10-100 THz range 
ubiquitously and naturally exist in their quantum ground state at 
room temperature. But unless they are individually addressed and resolved within
their coherence time, the ensemble average over unresolved vibrational modes precludes the observation of individual quanta. Despite this challenge, a Raman-active vibration featuring a specific form of internal nonlinearity was prepared in a squeezed state by optical excitation \cite{garrett1997}, while polarization-selective Raman interactions were used to observe two-photon interferences mediated by a vibrational mode \cite{lee2011}.
On the other hand, nano-fabricated mechanical oscillators can be susceptible to a universal radiation pressure interaction with light, especially with the intense fields stored in an optical cavity \cite{aspelmeyer2014}. 
However, their relatively low frequency (MHz-GHz) means that thermal energy
at room temperature is larger than the energy of a single vibrational quantum, making quantum state manipulation difficult or impossible under ambient conditions. 
Precise measurements of mechanical motion at room temperature have recently revealed quantum characteristics of the underlying radiation pressure interaction \cite{purdy2017,sudhir2017,Cripe19}. But it is only through deep cryogenic operation that quantum states of motion of nano-scale oscillators have been prepared in the last years \cite{wollman2015,hong2017a, riedinger2018, ockeloen-korppi2018,chu2018}. 
Thus, the quest to prepare quantum states of commonly available mechanical oscillators at ambient conditions remains largely open.

Here we prepare a non-classical state of an internal vibration of a diamond crystal at room temperature.
In a scheme inspired by the DLCZ protocol \cite{duan2001,galland2014}, a femtosecond laser pulse (the ``write'' pulse hereafter) first creates an excitation from the ambient motional ground state with a probability $p\ll 1$ (see Fig.~\ref{fig:intro}c). 
Detection of the emitted Stokes photon heralds the success of this step, while choice of the spectral window for detection fixes the specific vibrational mode of the  system under study.
To verify that only a single quantum of vibration was excited, a second laser pulse (the ``read'' pulse) retrieves it as an anti-Stokes photon. The probability of having two or more anti-Stokes photons, and therefore two or more quanta of vibrations, is obtained by performing a Hanbury-Brown-Twiss (HBT) intensity correlation measurement on the anti-Stokes photons conditioned on the heralding signal \cite{grangier1986,mandel1995}.
We observe sub-Poissonian statistics of the heralded vibrational state, a result consistent with having prepared the Fock state $\ket{1}$ of the vibrational mode (the first excited energy eigenstate).
Finally by changing the delay between the write and read pulses we
probe the decay of the single vibrational quantum, with $\approx 200$~fs resolution.

In contrast to previous quantum optics experiments on Raman-active vibrational modes that were restricted to molecular or crystal structures exhibiting particular polarization rules \cite{lee2011,lee2012,england2013,england2015,hou2016,fisher2016a,fisher2017} or vibrational nonlinearities \cite{garrett1997},
our technique is agnostic to these details, and can be employed on any Raman-active subject. 
It opens a plethora of opportunities to study vibrational quantum states and dynamics in other crystals and in molecules, and can readily be extended to create vibrational two-mode entangled states \cite{flayac2014} and test the violation of Bell inequalities \cite{vivoli2016,marinkovic2018} at room-temperature in a number of widely available systems.\\


\tocless\section{Theoretical description}

The Raman interaction between a single vibrational mode and an optical field leads to the creation of an anti-Stokes (Stokes) photon commensurate with the destruction (creation) of a vibrational quantum.
In our experiment the Raman interaction is driven by an optical field that may either be a ``write'' (superscript w) or 
``read'' (superscript r) pulse defined by the spatial and temporal mode of a mode-locked laser whose beam is focused onto the sample. 
These two incident fields are described by the annihilation operators $\hat{a}^\t{w,r}$.
The interaction leads to the generation of Stokes and anti-Stokes photons whose spatial mode is post-selected by projecting the focal spot onto the core of a single-mode optical fiber.
The Stokes (anti-Stokes) fields are modeled by annihilation 
operators $\op{a}_\t{S}^\t{w,r}$ ($\op{a}_\t{AS}^\t{w,r}$).
Due to conservation of energy and momentum in the Raman scattering process, the detection of these scattered 
fields as described above defines a single spatio-temporal mode of the vibration that is the subject of the experiment, 
and which we describe by its annihilation operator $\op{b}$. 
The Raman interaction is modeled by the Hamiltonian \cite{vonfoerster1971},

\begin{subequations}\label{interaction_Hamiltonian}
\begin{align}
	\op{H}^{\t{w}}_\t{int} &= i\hbar\left[ 
		G_\t{S}^\t{w}\, \hat{a}^\t{w} \hat{b}^\dagger (\hat{a}_\t{S}^\t{w})^\dag +
		G_\t{AS}^\t{w}\, \hat{a}^\t{w} \hat{b}(\hat{a}_\t{AS}^\t{w})^\dagger \right] + \t{h.c.} \\ 
	\op{H}^{\t{r}}_\t{int} &= i\hbar\left[ 
		G_\t{S}^\t{r}\, \hat{a}^\t{r} \hat{b}^\dagger (\hat{a}_\t{S}^\t{r})^\dag +
		G_\t{AS}^\t{r}\, \hat{a}^\t{r} \hat{b}(\hat{a}_\t{AS}^\t{r})^\dagger \right] + \t{h.c.},
\end{align}
\end{subequations}
where the coupling rates $G_\t{S}^\t{w,r}$ and $G_\t{AS}^\t{w,r}$ relate to the Raman 
activity of the vibrational mode.

None of the four processes described by \cref{interaction_Hamiltonian} is resonant since
we work at photon energies well below the band gap of diamond (5.47~eV). 
However, because we spectrally filter and detect only the two modes $a_\t{S}^\t{w}$ and $a_\t{AS}^\t{r}$ all essential results of our measurements can be described by the Hamiltonian
\begin{subequations}\label{reduced_Hamiltonian}
\begin{align}
	\op{H}^{\t{w}}_\t{int} &= i\hbar g_\t{S}^\t{w}\, \hat{b}^\dagger \hat{a}_\t{S}^\dag  + \t{h.c.}  \label{write_Hamiltonian} \\
	\op{H}^{\t{r}}_\t{int} &= i\hbar	g_\t{AS}^\t{r}\, \hat{b} \, \hat{a}_\t{AS}^\dagger  + \t{h.c.}, \label{read_Hamiltonian}
\end{align}
\end{subequations}
where we defined the effective coupling rate, $g_\t{S,AS}^\t{w,r} \equiv G_\t{S,AS}^\t{w,r} \sqrt{n_p^\t{w,r}}$, corresponding to a classical excitation for the write/read pulses with $n_p^\t{w,r}$ photons per pulses, and used the shorter notation $\hat{a}_\t{S} \equiv \hat{a}_\t{S}^\t{w}$, $\hat{a}_\t{AS} \equiv \hat{a}_\t{AS}^\t{r}$.
In fact, this scenario is equivalent to the linearized radiation-pressure 
interaction \cite{aspelmeyer2014}, or Raman processes in atomic ensembles \cite{duan2001}, where
an optical cavity (in the former instance) or electronic resonance (in the latter) suppresses non-resonant
terms.

Note that the Hamiltonian in \cref{reduced_Hamiltonian} neglects higher order interactions, in particular the creation of
correlated Stokes/anti-Stokes pairs during a single pulse (write or read) via phonon-assisted four-wave 
mixing \cite{bloembergen1965,kasperczyk2015,parra-murillo2016,schmidt2016,kasperczyk2015,kasperczyk2016,saraiva2017}.
In our experiment, the photon flux due to this higher-order interaction constitutes an effective background noise \cite{anderson2018}. 
Other extraneous sources of photons, such as from residual 
$\chi^{(3)}$ nonlinearities, or fluorescence, also lead to an excess background noise.
The spontaneous Stokes signal scales linearly with laser power since it is proportional to $n_p^\t{w}(\bar{n}+1)\approx n_p^\t{w}$, where $\bar{n}\ll 1$ is the average occupation of the vibrational mode. 
In our experiment, the spontaneous Raman signal is much stronger than any of the parasitic processes described above, so that the fidelity of the heralded state is only marginally affected by noise. 
However, the anti-Stokes signal, being proportional to $n_p^\t{r}\bar{n}$, 
is significantly weaker, so that noise cannot be neglected in this case. The measured normalized probability of detecting two 
anti-Stokes photon should therefore be considered as an upper-bound for the corresponding probability of having two vibrational quanta. 
In the calculations presented below, the noise terms are accounted for in order to faithfully describe the experiment.\\

\tocless\subsection{Write operation}

The first step in an iteration of the experiment is the excitation of the vibrational mode by 
a write pulse. The resulting dynamics of the vibrational and Stokes modes is governed by 
$H_\t{int}^\t{w}= i\hbar \, g_\t{S}^\t{w}\, \hat{b}^\dag \hat{a}_\t{S}^\dag+ \t{h.c.}$, 
which has the form of a two-mode squeezing interaction, and leads to the creation of maximally correlated 
pairs of vibrational and Stokes excitations. 
When both modes are initially in the vacuum state $\ket{\text{vac}}$, the final state after the write 
pulse is \cite{mandel1995},
\begin{equation}\label{2mode_squeezed}
	\ket{\Psi}_\t{S,b}= \sqrt{1-p} \sum_{n=0}^\infty \sqrt{p^n}\ket{n,n}_\t{S,b}
\end{equation}
where $\ket{n,n}_\t{S,b}\equiv \frac{\left(\hat{b}^\dag\hat{a}_\t{S}^{\dag}\right)^n}{n!} \ket{\text{vac}}$. 
For the simple situation of a constant interaction switched on for a duration $T_\t{w}$, the probability of
exciting the state $\ket{1,1}$ is given by, $p = \tanh^2(g_\t{S}^\t{w} T_\t{w})$. For realistic laser pulse shapes, in the linear regime, $T_\t{w}$ is the effective interaction time defined by the equivalent square pulse that carries the same pulse energy. 
Ideally, when at least one Stokes photon is detected, the (conditional) state of the vibrational mode 
becomes (see Supplementary information), $\rho_\t{b\vert S} \approx \ket{1}\bra{1}+p\ket{2}\bra{2}$, in the 
limit where $p\ll 1$; crucially, the vacuum component in $\ket{\Psi}_\t{S,b}$ has been eliminated based 
on the presence of a Stokes photon.
Dark noise in real photodetectors (modeled as a probability $\pi_0$ per pulse) prevents 
unambiguous discrimination of the vacuum contribution. 
However, when the total Stokes signal is larger than the dark noise, it can
be shown (see Supplementary information) that the resulting conditional state,
\begin{equation}\label{cond_state}
	\hat{\rho}_\t{b\vert S} \approx \frac{\pi_0}{2\eta p}\ket{0}\bra{0} + \ket{1}\bra{1} + p\ket{2}\bra{2},
\end{equation}
is dominated by the contribution from the pure Fock state $\ket{1}$. Here $\eta$ is the detection efficiency of the Stokes
field, and $\eta p$ is the Stokes detection probability. The signal-to-noise ratio in the Stokes photodetector $\eta p/\pi_0$ is larger than $10^4$ in our experiment.\\

\tocless\subsection{Read operation}

Once the Stokes photon is detected, a second pulse -- the read pulse -- 
is used to retrieve the (conditional) state of the vibrational mode.
The dynamics of the anti-Stokes field induced by the read pulse is described by the Hamiltonian 
$H_\t{int}^\t{r}= i\hbar \, g_\t{AS}^\t{r}\, \hat{b} \, \hat{a}_\t{AS}^\dag + \t{h.c.} $, 
which represents a beam splitter interaction between the anti-Stokes and vibrational modes. 
For an effective interaction time $T_\t{r}$, the state of the emitted anti-Stokes field is governed by the 
input-output relation, 
\begin{equation}\label{read_input_output}
	\hat{a}_\t{AS,out}(T_\t{r}) = \cos(\theta)\hat{a}_\t{AS,in} + \sin(\theta) \hat{b},
\end{equation}
where, $\theta \equiv g_\t{AS}^\t{r} T_\t{r}$. 
The mode $\hat{a}_\t{AS,in}$ describes the input anti-Stokes mode (before the read pulse), 
which is in the vacuum state. 

The crucial aspect of the read operation is that the emitted anti-Stokes field faithfully reflects the number
statistics of the vibrational mode. In fact, from \cref{read_input_output}, we find that
$\langle :( \hat{a}_\t{AS}^{\dag} \hat{a}_\t{AS})^n :\rangle 
	= \sin^{2n}(\theta) \langle : ( \hat{b}^{\dag} \hat{b} )^n: \rangle$,
for any integer $n~\geq~1$ (here $::$ denotes normal ordering). 
This relation expresses the fact that photon counting is insensitive to vacuum noise (in stark contrast 
to linear detection of the field \cite{weinstein2014}), so that normalized moments of the photon
number of the anti-Stokes field faithfully represent the statistics of the vibrational mode excitation number.

Note that although the interaction between the optical pulses and the vibrational mode
is linear (\cref{reduced_Hamiltonian}), the nonlinearity provided by the single-photon detection following the
write and read pulses renders the full measurement process effectively nonlinear \cite{klm2001}.
It is this single-photon nonlinearity that lies at the conceptual heart of our protocol.\\

\tocless\subsection{Statistics of the heralded intensity correlation} 

Performing the two operations presented above enables the preparation and unambiguous characterization of a vibrational Fock state.
Consider the joint state $\ket{\Psi}_\t{S,b,AS}$ for the Stokes field, the vibration, and the anti-Stokes field. 
A heralded coincidence event occurs when a Stokes photon is detected at time $t=0$ , followed, after a time $t$, by coincident detection of a pair of anti-Stokes photons.
This heralded coincidence event is represented by the measurement map, 
\begin{equation*}
	\ket{\Psi}_\t{S,b,AS} \mapsto \hat{d}_2 (t) \hat{d}_1 (t) \hat{a}_\t{S}(0) \ket{\Psi}_\t{S,b,AS},	
\end{equation*}
where $\hat{d}_{1,2}$ are the operators denoting the anti-Stokes field at the two output of a 50/50 beam-splitter (see \cref{fig:intro}c).
The probability of this triple coincidence defines the conditional intensity correlation, and 
is thus proportional to
$\mean{\hat{a}_\t{S}^{\dagger} (0) \hat{d}_1^\dagger(t)\hat{d}_2^\dagger(t) \hat{d}_2(t) 
\hat{d}_1(t) \hat{a}_\t{S}(0)}$; here we have used the linearity of quantum mechanics to extend the definition to mixed states as well.
Suitably normalizing the expression \cite{glauber1963a} 
allows us to define the conditional intensity correlation,
\begin{equation*}
	g_{\t{AS}\vert\t{S}}^{(2)}(t) \equiv
	\frac{ 
		\mean{\hat{a}_\t{S}^{\dagger} (0) \hat{d}_1^\dagger (t) \hat{d}_2^\dagger (t) \hat{d}_2(t) 
		\hat{d}_1(t) \hat{a}_\t{S}(0)}
	}{
		\prod_{i}\left[  
			\mean{\hat{a}_\t{S}^{\dagger} (0)\hat{d}_i^\dagger (t)\hat{d}_i(t)a_\t{S}(0)}
			\mean{\hat{a}_\t{S}^{\dagger} (0) \hat{a}_\t{S}(0)}^{-1/2}
		\right]
	}.
\end{equation*}
The fields $\hat{d}_{1,2}$ whose intensity cross-correlation is measured can be expressed in terms of the 
anti-Stokes field $\hat{a}_\t{AS}$, which in turn
can be expressed in terms of the vibration (via \cref{read_input_output}); since intensity correlations do not
respond to the vacuum, the open port of the beam-splitter used in the intensity correlation measurement plays no role, and
the conditional correlation above can be written as,
\begin{equation}\label{g2_cond_threefold}
		g_{\t{AS}\vert\t{S}}^{(2)}(t) =
	\frac{ 
		\mean{\hat{a}_\t{S}^{\dagger} (0) \hat{b}^\dagger (t) \hat{b}^\dagger (t) \hat{b}(t) 
		\hat{b}(t) \hat{a}_\t{S}(0)}
	}{
		\mean{\hat{a}_\t{S}^{\dagger} (0) \hat{a}_\t{S}(0)}^{-1}\mean{\hat{a}_\t{S}^{\dagger} (0)
		\hat{b}^\dagger (t)\hat{b}(t)\hat{a}_\t{S}(0)}^2
	}.
\end{equation}
After the detection of a Stokes photon (i.e. for $t > 0$), the state of the vibrational mode (\cref{cond_state}) 
has disentangled from that of the Stokes mode, so that expectation values of products of operators in their 
joint state factorize into products of expectation values; thus,
\begin{equation}\label{g2_cond_g2b}
	g_\t{AS\vert S}^{(2)}(t >0) = \frac{ 
		\mean{\hat{b}^\dagger (t) \hat{b}^\dagger (t) \hat{b}(t) \hat{b}(t)}_\t{\vert S} }
		{ \mean{\hat{b}^\dagger (t)\hat{b}(t)}_\t{\vert S}^2} = g^{(2)}_{b\vert S}.
\end{equation}
That is, the conditional intensity correlation of the anti-Stokes field gives the intensity correlation of
the vibrational mode.
Immediately after the write pulse, i.e. at $t =0$, and in the limit of a small probability $p$ of exciting a 
vibrational Fock state, explicit evaluation of 
\cref{g2_cond_threefold} on the state $\ket{\Psi}_\t{S,b}$ of \cref{2mode_squeezed} yields
\begin{equation}\label{g2cond_value_QM}
	g^{(2)}_{\t{AS}\vert\t{S}}(0) \approx \frac{4P(2,2)}{P(1,1)}=4 p
\end{equation}
where $P(n,n)=  \bra{n,n} \left( \ket{\Psi}_\t{S,b} \bra{\Psi} \right) \ket{n,n}  =p^n$
is the probability of finding $n$ pairs of excitations in the vibrational mode and  Stokes field upon a projective measurement on state (\ref{2mode_squeezed}) in the Fock state basis.

In our experiment we measure the number of events $N_{d_1,d_2,a_\t{S}}$ where photons were detected simultanously in modes $d_1, d_2$, and $a_\t{S}$ (i.e. triple coincidence), and normalize it to the product of the number of events
$N_{d_i, a_\t{S}}$ ($i=1,2$) where photons are detected simultaneously in the Stokes mode and one of the anti-Stokes
detectors (i.e. a two-fold coincidence); we thus measure \cite{grangier1986},
\begin{equation}\label{alpha_num}
	\alpha \equiv \frac{ N_{d_1,d_2,a_\t{S}} N_{a_\t{S}} }{N_{d_1,a_\t{S}}N_{d_2,a_\t{S}}}.
\end{equation}
It is important to note that, in general, $\alpha$ is not equivalent to $g^{(2)}_{\t{AS}\vert\t{S}}$ in 
\cref{g2cond_value_QM}. 
More precisely, if the detection efficiency of the herald mode $a_\t{S}$ is $0<\eta\leq 1$ (which we model 
as a beam splitter with transmittance $\eta$ placed before the detector) we find (see Supplementary information)
\begin{equation}\label{alpha_value}
	\alpha \approx (4-2 \eta)\frac{P(2,2)}{P(1,1)}=(4-2 \eta)p.
\end{equation} 
Thus, in the limit of low detection efficiency of Stokes photons (i.e. $\eta\ll 1$) we have that 
$\alpha = g^{(2)}_{\t{AS}\vert \t{S}}$. 
In the experiment, $\eta \approx 10\%$, so that $\alpha \approx g^{(2)}_{\t{AS}\vert \t{S}}$ (to within 5~\%).
  

\tocless\section{Experimental Realization} 

\tocless\subsection{Setup and measurement procedure}

Our experimental setup is an upgraded version of that presented in ref. \cite{anderson2018}. Two
synchronized laser pulse trains at 810~nm and 695~nm of duration $\Delta t \approx 100$~fs  are
produced by a Ti:Sa oscillator (Tsunami, Spectra Physics, 80 MHz repetition rate) and a synchronously pumped
frequency-doubled optical parametric oscillator (OPO-X fs, APE Berlin), respectively. The write
pulses are provided by the OPO, while the Ti:Sa provides the read pulses, which are passed
through a delay line before being overlapped with the OPO output on a dichroic mirror. The sample is a
synthetic diamond crystal ($\sim 300$~$\mu$m thick, from LakeDiamond) cut along the (100) crystal
axis and is probed in transmission using two microscope objectives (numerical aperture 0.8 and 0.9).
The laser light is blocked using long-pass and short-pass tunable interference filters (Semrock),
leaving only a spectral window of transmission for the Stokes signal from the write pulse (mode $a_\t{S} $) and the 
anti-Stokes signal from the read pulse (mode $a_\t{AS} $). 
The transmission is collected in a single mode fiber (for spatial mode filtering) 
and then the two signals are separated with a tunable long-pass filter used as a dichroic mirror. 
After an additional band-pass filter (see Supplementary Information, Section VI) each signal is coupled into a multi-mode fiber; subsequently,
the Stokes field is sent to a single photon counting module (SPCM, Excelitas), while the anti-Stokes
field is split at a 50:50 fiber beam-splitter and directed onto two SPCM's. The three
SPCM's are then connected to a coincidence counter (PicoQuant TimeHarp 260).

We only record the coincidence events where a click in one of the anti-Stokes channels was preceded by
a click in the Stokes channel. This allows us to find heralded coincidence events (within the same laser repetition),
as well as to build a delay histogram using the Stokes channel as the start and either of the anti-Stokes 
channels as the stop. 
These start-stop histograms are used to compute the Stokes -- anti-Stokes intensity cross-correlation as explained in \cite{anderson2018}. 
Therefore all relevant coincidences required to estimate $\alpha$ are available.

\begin{figure}[t!]
	\centering
	\includegraphics[width=\columnwidth]{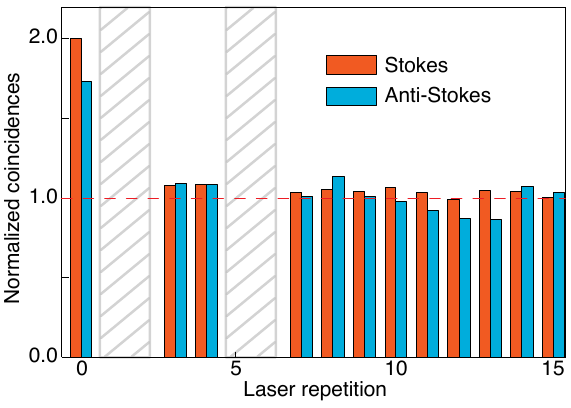}
	\caption{\label{fig:aSaS} 
	\textbf{Unconditional Stokes and anti-Stokes correlations.}
	Two-photon coincidence histograms of the Stokes field $\op{a}_\t{S}$ (write pulse energy 60 pJ, acquisition time 10~min) and anti-Stokes field $\op{a}_\t{AS}$ (read pulse energy 372 pJ, acquisition time 60~min). For the anti-Stokes field the coincidences are recorded between the detectors measuring $\hat{d}_1$ and $\hat{d}_2$ as in Fig.~\ref{fig:intro}b, while the write pulse is blocked. For the measurement of the Stokes field, a beam splitter is added in the path of mode 
	$\op{a}_\t{S}$. 
	The start-stop delay (horizontal axis) is scaled in multiples of the repetition period $\approx 12.5$~ns. After normalizing by the average number of accidental coincidences (the peaks non-zero start-stop) the value at zero time delay represents the intensity correlation of the Stokes and anti-Stokes fields, namely $g_\t{S}^{(2)}(0)=2.0 \pm 0.1$ and $g_\t{AS}^{(2)}(0)=1.73 \pm 0.11$.
	The hatched region in gray, omitted in the analysis, are affected by spurious coincidences due to cross-talk between the two detectors, which arise when hot-carrier-induced light emission from one detector \cite{lacaita1993} is received by the other detector. 
}
\end{figure}
~\\

\tocless\subsection{Ambient thermal state} 

We start by verifying that following the write pulse the Stokes field is well described by the state of \cref{2mode_squeezed}: when marginalized over the state of the vibrational mode, the Stokes field is thermal.
Indeed, we find in Fig.~\ref{fig:aSaS} (red bars) that the intensity correlation function of the Stokes field at zero time delay is $g^{(2)}_\t{S}(0) = 2$, {as expected for a single mode thermal state}. Similarly,
in the absence of the write operation, the anti-Stokes signal should reflect the thermal statistics of the vibrational mode. 
To check this, we measure the (unconditional) intensity correlation of the anti-Stokes mode, 
shown in Fig.~\ref{fig:aSaS} (blue bars). 
The value of $g^{(2)}_\t{AS}(0) = 1.73 \pm 0.11$, is slightly lower than the expected value of 2 for a single mode thermal state, 
but higher than the value $1+\frac{1}{N}$ for a thermal state of $N>1$ modes \cite{sekatski2012}. 
We attribute this discrepancy to noise in the anti-Stokes channel coming from degenerate four-wave mixing in the sample 
(which includes the second-order Stokes--anti-Stokes process \cite{parra-murillo2016} discussed earlier).  
We thus confirm a single vibrational mode in an ambient thermal state, and that the result of the write operation is well described by the two-mode squeezed state of \cref{2mode_squeezed}.

\begin{figure}[t!]
	\centering
	\includegraphics[width=\columnwidth]{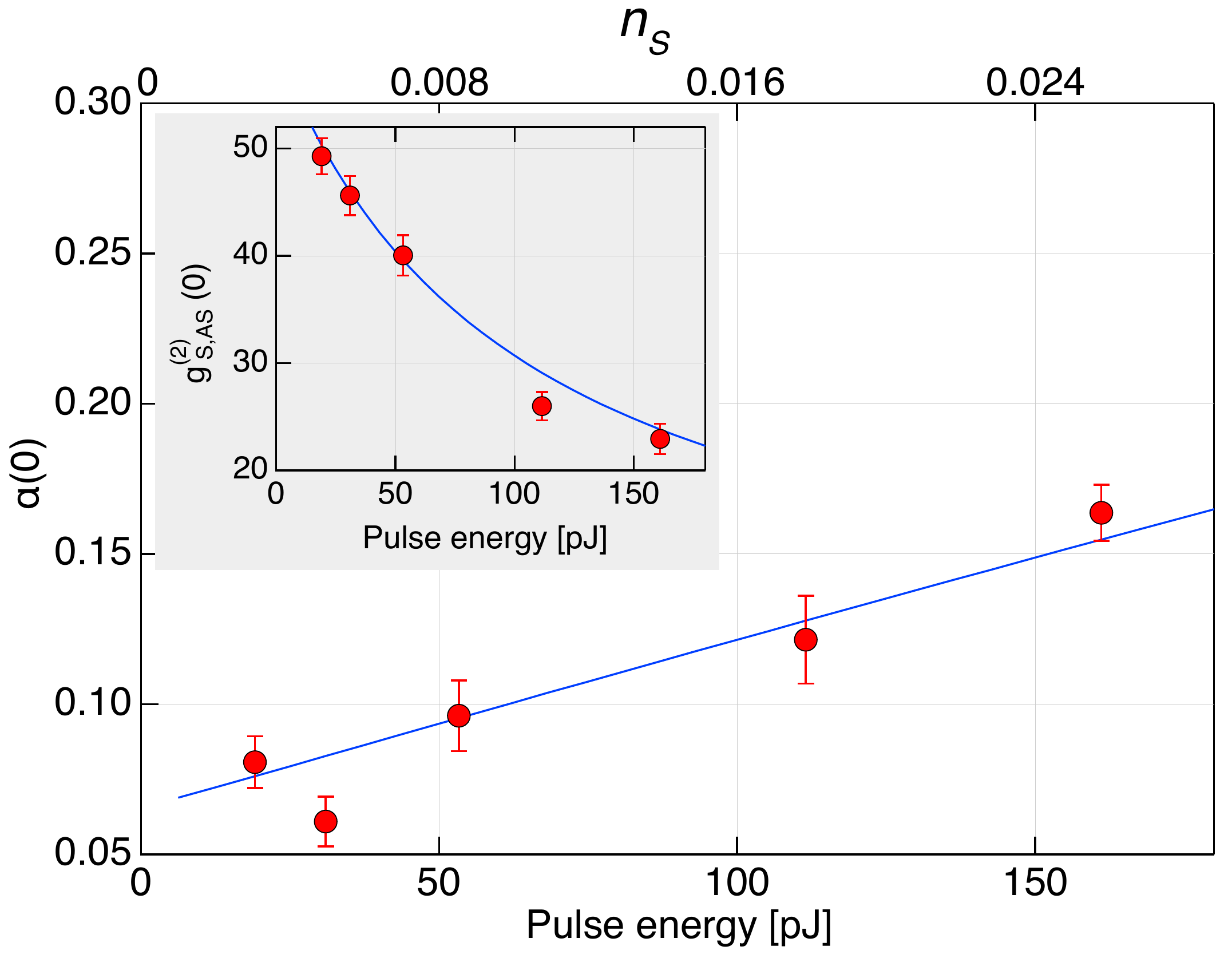}
	\caption{\label{fig:PowDep} 
	\textbf{Sub-Poissonian anti-Stokes statistics.}
	Dependence of the heralded vibrational statistics on the write pulse energy and on the corresponding estimated probability of creating at least one Stokes photon per pulse, $n_\t{S} = \frac{p}{1-p}$. The normalized Stokes--anti-Stokes correlations are shown as an inset. The write--read pulse delay is fixed at zero and the read pulse energy at $322$~pJ. Statistical error bars are obtained from the square root of the total number of events. 
	Blue lines are models (see Supplementary Information), using the estimated detection efficiency (10~$\%$) and a relative efficiency of the read process (relative to the Stokes emission cross-section) of 30~$\%$ as the only two adjustable parameters (common to both panels).
}\end{figure}
~\\

\tocless\subsection{Fock state prepartion}

In order to prepare the vibrational mode in a Fock state, we send a write pulse and herald the success of this 
operation by detecting a Stokes photon. 
When a subsequent read pulse retrieves the vibrational state, and is
subjected to intensity correlation measurements, we find that $\alpha(0) \approx g_\t{AS\vert S}^{(2)}(0) < 1$, 
as shown in \cref{fig:PowDep} (main panel). Thus, the conditional anti-Stokes field
exhibits sub-Poissonian statistics. 
But since we know that the anti-Stokes field is faithful to the vibrational state, and specifically that
$g^{(2)}_\t{AS\vert S} = g^{(2)}_\t{b|S}$, we are able to conclude that the vibrational mode exhibits sub-Poissonian 
number statistics.
From the value of $\alpha(0)\approx 0.06$ at the lowest powers of the write pulse and the known detection efficiency of the
Stokes field $\eta\approx 10\%$, our theoretical model allows us to estimate the probability of having excited 
the Fock state $\ket{1}$ to be (\cref{alpha_value}), $1-p \approx 98.5\%$.

With increasing power of the write pulse, mixtures of states higher up in the Fock ladder are excited.
As shown in Fig.~\ref{fig:PowDep}, the sub-Poissonian character of $\alpha$ decreases with increasing pump power, 
as expected from the simple model $\alpha(t=0)\propto p = \tanh^2(g_\t{S}^\t{w} T_\t{w}) \approx 
(g_\t{S}^\t{w} T_\t{w})^2 \propto n_p^\t{w}$, where $n_p^\t{w}$ is the number of photons per write pulse.
in tandem, the Stokes--anti-Stokes correlation reduces as $1/n_p^\t{w}$ (Fig.~\ref{fig:PowDep} inset).
These trends are consistent with increasing probability of exciting two or more
vibrational quanta \cite{friberg1985,sekatski2012} (see Supplementary Information).\\

\begin{figure}[t!]
	\centering
	\includegraphics[width=\columnwidth]{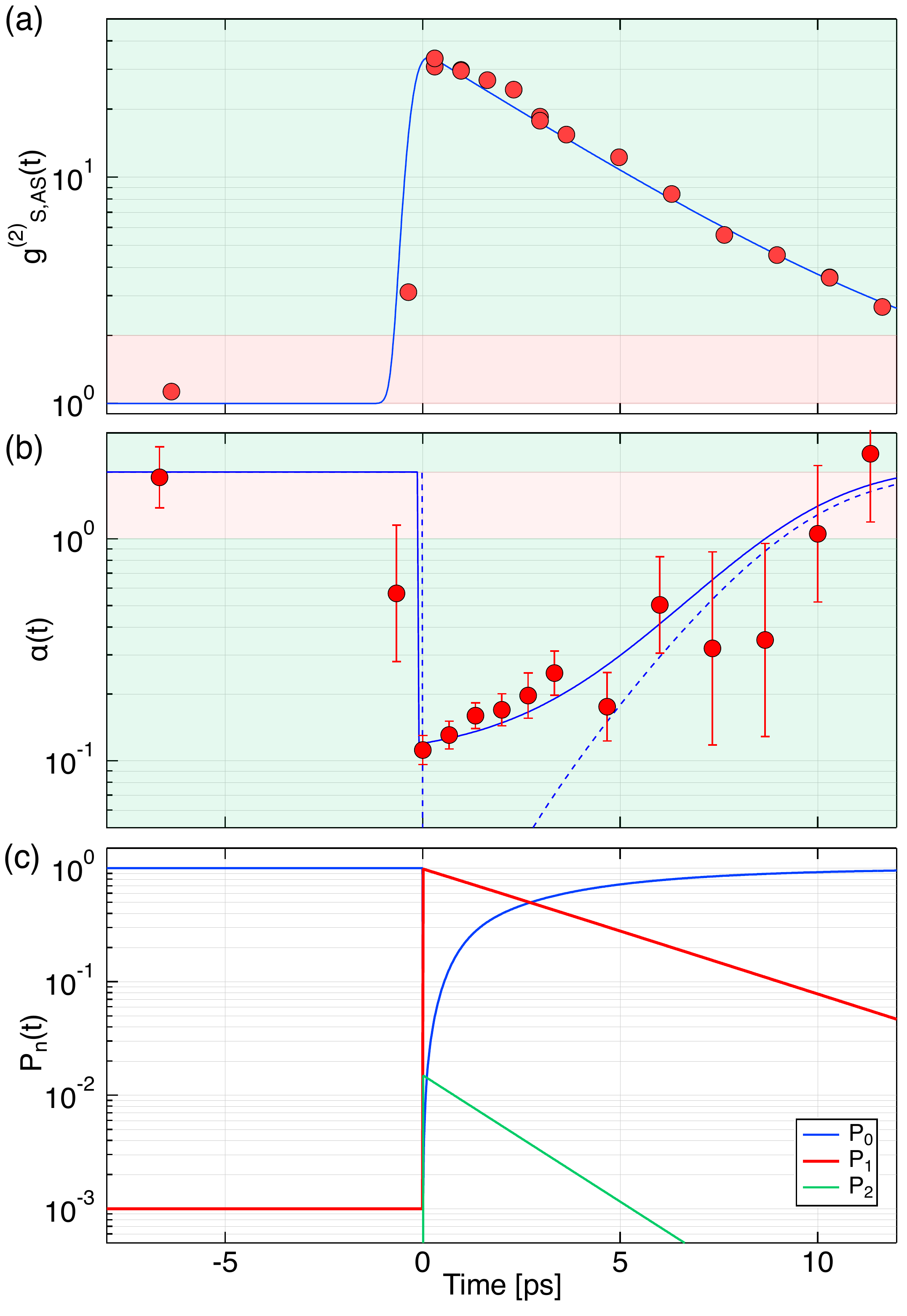}
	\caption{\label{fig:TimeDep} 
	\textbf{Decay of vibrational Fock state.}
	Measured Stokes--anti-Stokes correlations (a), and
	heralded vibrational mode intensity correlation (b) as a function of write--read delay.
	The measurements (full circles) are taken with a pulse energy of $62$~pJ in the write pulse and $409$~pJ in the read pulse, with an acquisition time of $60$~min for all points except the one at $-6.3$~ps, which was acquired over $8$~hours.
	Statistical error bars are obtained from the square root of the total number of events detected in each case. 
	Full circles are measured data and blue lines are from the model, see main text and eq.~(\ref{fit_alpha}).
	The prediction from the model without the added noise on the anti-Stokes detectors is shown with a dashed line in (b).
	Green bands in panels (a,b) show the region where a non-classical model is required to explain the
	observations; red bands indicate where a classical model suffices.
	(c) Fock state distribution of the conditional vibrational mode inferred from the data via our model.
}\end{figure}


\tocless \subsection{Fock state dynamics}

The decay of the excited vibrational Fock state can be probed by allowing it to evolve freely after the write pulse. 
In the experiment, we do this by employing a variable optical path length to impose a time delay $t$ between the write 
and read pulses.
\Cref{fig:TimeDep} summarizes the observed time-dependence of the excited Fock state. 
When the write and read pulses overlap ($t = 0$), we observe strong Stokes--anti-Stokes number correlation $g^{(2)}_\t{S,AS}(0) \approx 30$, consistent with the generation of a Stokes-vibration two-mode squeezed state (\cref{2mode_squeezed}).

Simultaneously, $\alpha$, which reflects the intensity correlation of the conditioned vibrational state 
(\cref{cond_state}), indicates sub-Poissonian statistics of the vibrational mode, with $\alpha(t=0) \approx 0.11 < 1$. 
Conditional on the detection of a Stokes photon, the vibrational mode is thus faithfully prepared in the Fock state $\ket{1}$.

Subsequent iterations of the experiment probe the vibrational state after a controlled time delay. 
\Cref{fig:TimeDep}a shows the decay of the Stokes--anti-Stokes correlation. 
The initial value $g^{(2)}_\t{S,AS}(0)$ quantifies the degree to which the Stokes field and vibrational mode are correlated by the write operation; at later times $t>0$, after the Stokes field is detected, $g^{(2)}_\t{S,AS}(t) \propto \langle \op{b}^\dagger (t) \op{b}(t)\rangle_\t{\vert S} $, so that the data in Fig.~\ref{fig:TimeDep}a, 
in conjunction with a model (shown in blue, consisting of the ideal prediction convolved with the known
instrument response), allows us to infer the decay rate $\tau_m = 3.9\pm 0.3\, \t{ps}$ (bounds for 95\% confidence). This value is consistent with the previously reported vibrational lifetime of $~3.6$~ps \cite{lee2012}.

In parallel, as shown in \cref{fig:TimeDep}b, $\alpha(t)$ mirrors this evolution, starting at $\alpha(t<0) = 1.9 \pm 0.6$ (thermal state), dropping to $\alpha(0) = 0.11 \pm 0.01$ at zero delay (Fock state $\ket{1}$), and then returning toward its equilibrium value as the prepared vibrational Fock state thermalizes with its environment.
(The larger uncertainty in the data at long and at negative delays is due to the reduced rate of coincidences, because of the small thermal occupancy, $\bar{n} \approx 1.5 \pw{-3}$, of the vibrational mode.)
This behaviour is captured by a simple model (shown as a blue line in Fig.~\ref{fig:TimeDep}b), based on the fact that $\alpha(t)$ is the intensity correlation of the vibrational mode (\cref{g2_cond_g2b}) -- which can be calculated using an open quantum system model for the vibrational mode -- together with a contribution from background noise in the anti-Stokes field 
(see Supplementary Information),
\begin{equation}\label{fit_alpha}
	\alpha(t) \approx \frac{2}{P_1(0)}\left[1-\frac{1}{(1+\bar{n}(e^{t/\tau_m}-1))^2} \right] + \alpha_0,
\end{equation}
where $P_1(0)\approx 0.985$ is the probability of having created the vibrational Fock state $\ket{1}$ conditioned
upon the detection of a Stokes photon, and $\alpha_0 = 2 \times 0.04$ corresponds to twice the 
anti-Stokes noise-to-signal ratio conditioned on Stokes detection (i.e., $\alpha_0/2 \approx 1/g^{(2)}_\t{S,AS}(0)$, see Supplementary Information).
The decay of sub-Poissonian statistics of the vibrational mode is consistent with the decay of the 
Fock state $\ket{1}$ to the ground state $\ket{0}$.

The measured decay of the conditional intensity correlation, in conjunction with a decoherence model, can be 
used to extract the number distribution $P_n(t)$ of the conditional vibrational state (see Supplementary Information) -- 
the probability to find the vibrational mode in the Fock state $\ket{n}$ at time $t$ after a click on 
the Stokes detector.
This projection is plotted in Fig.~\ref{fig:TimeDep}c. 
Noteworthy is the high purity of the conditional vibrational state with respect to the Fock state $\ket{n=1}$, 
since its normalized second order correlation is $\frac{2 P_2}{P_1^2}\approx0.02$. This is much lower than the 
measured parameter $\alpha\approx0.1$ (because the background noise impinging on the anti-Stokes detectors 
affects $\alpha$) and highlights the potential of the technique to produce high-purity single phonons. \\


\tocless \section{Conclusion} 

We have demonstrated for the first time that a high-frequency Raman-active vibrational mode can be prepared in its $n=1$ Fock state at room temperature. 
Heralded intensity correlation measurements confirm the sub-Poissonian statistics of the conditional vibrational state.
We further probed the decay of the vibrational Fock state, akin to similar measurements on microwave photons \cite{wang2008,brune2008}. 

This research opens a door to the study of quantum effects in the vibrational dynamics of Raman-active modes in immobilized molecules \cite{yampolsky2014}, liquids, gases \cite{bustard2013,bustard2015,bustard2016} and solid-state systems. 
Vibrational states in Raman-active solid-state systems at room temperature may even be viable candidates for quantum technology if the coherence time and readout efficiency can be improved. 
Coherence of the longitudinal optical phonon modes of diamond is known to be limited by decay through the so-called
Klemens channel \cite{klemens1966, debernardi1995, liu2000}. Proposals to close this pathway include the creation of a phononic band gap at the atomic scale obtained by growing ${}^{12}C - {}^{13}$C super lattices \cite{lee2012}.
Readout efficiency and heralding rate could be improved by coupling the Raman-active system to small-mode volume optical or plasmonic cavities \cite{roelli2016,benz2016a,lombardi2018}, or by employing resonant Raman scattering \cite{jorio2014,jiang2018}. Molecular systems are particularly promising for extending the vibrational lifetime \cite{chen2019}. 
With these improvements, vibrational modes may be used as a source of high-purity on-demand (anti-Stokes) photons,
or as a buffer memory to produce heralded single photons with an arbitrary choice of the herald and signal wavelengths and/or bandwidths, or even heralded frequency conversion at the single photon level.


\clearpage
\begin{center}
	{\large \bfseries Supplementary Information}
\end{center}

\tableofcontents

\appendix

\section{Initial state of vibrational mode}\label{app:initial}

Immediately after the write pulse Raman interaction with the sample the joint state of the Stokes and vibrational modes is,
\begin{equation*}
	\ket{\Psi_\t{S,b}} = \sqrt{1-p} \sum_{n=0}^\infty \sqrt{p^n} \ket{n_\t{S}=n,n_\t{b}=n},
\end{equation*}
where $p$ is the probability of exciting at least on pair of Stokes - vibrational quanta.
The probability of detecting exactly one Stokes photon through an ideal photon number-resolving projective measurement would be $\frac{p}{1-p}$.

If we now consider ideal photodetection (unit efficiency and no background noise) with a ``click" (or ``bucket'') detector, which does not resolve photon number, the measurement is appropriately modeled by the positive-operator valued measure (POVM)
\begin{equation}\label{eq:PIideal}
	\op{\Pi} \equiv \sum_{n_\t{S}=1}^\infty \ket{n_\t{S}} \bra{n_\t{S}}
		= \mathbf{1}-\ket{0_\t{S}} \bra{0_\t{S}},
\end{equation}
which corresponds to the proposition ``at least one Stokes photon is detected''. 
 Note that this POVM is suggestive of the notion that detecting at 
least one Stokes photon ``projects out'' the vacuum contribution of the state $\ket{\Psi_\t{S,b}}$.
To establish this we compute the vibrational state after a detection event (i.e. a detector ``click"):
\begin{equation}\label{rhoCond}
\begin{split}
	\op{\rho}_\t{b\vert S} &\propto \t{Tr}_\t{S}\left[ 
		\op{\Pi} \ket{\Psi_\t{S,b}}\, \bra{\Psi_\t{S,b}} \op{\Pi}^\dagger \right] \\
	&= \sum_{k_\t{S}} \bra{k_\t{S}} \op{\Pi} \ket{\Psi_\t{S,b}}
		\bra{\Psi_\t{S,b}} \op{\Pi}^\dagger \ket{k_\t{S}} \\
	&\propto \sum_{k_\t{S},m_\t{S},n_\t{b}} \sqrt{p^{n+m}} \left[
		\bra{k_\t{S}} \op{\Pi} \ket{m_\t{S}} \bra{m_\t{S}} \op{\Pi}^\dagger \ket{k_\t{S}}\right]
		\ket{n_\t{b}} \bra{n_\t{b}}.
\end{split}
\end{equation}
The matrix elements (terms in the bracket) required to progress further can be computed using \cref{eq:PIideal}:
\begin{equation*}
	\bra{k} \op{\Pi} \ket{n}_\t{S} = \delta_{k,n} - \delta_{k,0}\delta_{n,0}.
\end{equation*}
Inserting this back and evaluating the sum gives,
\begin{equation*}
	\op{\rho}_\t{b\vert S} \propto \sum_{n=1}^\infty p^{n} \ket{n} \bra{n}.
\end{equation*}
which upon normalization gives the vibrational state conditioned on ideal photodetection of the Stokes field,
\begin{equation*}
	\op{\rho}_\t{b\vert S} = \frac{1-p}{p} \sum_{n=1}^\infty p^{n} \ket{n} \bra{n}
		\approx \ket{1} \bra{1} + p\ket{2} \bra{2}.
\end{equation*}
(The last approximation is valid to first order in $p$.)

\subsection{Case of non-ideal Stokes detection}\label{app:initial:real}

Photodetectors however do not respond with unit efficiency to the presence of photons, nor are they free of noise. Such non-ideal photodetectors can be modeled by the POVM \cite{sekatski2012},
\begin{equation*}
	\op{\Pi} \equiv \mathbf{1}-(1-\pi_0)(1-\eta)^{\op{a}_\t{S}^\dagger \op{a}_\t{S}},
\end{equation*}
where $\pi_0$ is the dark count probability, and $\eta$ is the detection efficiency.
The matrix elements of this operator on the Stokes photon states, 
\begin{equation*}
	\bra{k} \op{\Pi} \ket{n}_\t{S} =
		\delta_{k,n}\left[1-(1-\pi_0)(1-\eta)^n \right],
\end{equation*}
when inserted into \cref{rhoCond} gives,
\begin{equation*}
\begin{split}
	\op{\rho}_\t{b\vert S} \propto & \pi_0^2 \ket{0} \bra{0} \\
		&\, + \sum_{n=1}^\infty p^n \left[1-(1-\pi_0)(1-\eta)^n \right]^2 \ket{n} \bra{n}.
\end{split}
\end{equation*}
Since we are interested in the case where the probability of exciting one or more vibrational quanta is low ($p \ll 1$),
we only consider the first 3 terms of this state; further, we operate in the regime of low detection
efficiency ($\eta \ll 1$) for reasons detailed in the main text, and with a low dark-count probability ($\pi_0 \ll 1$).
Under these conditions, 
\begin{equation*}
\begin{split}
	\op{\rho}_\t{b\vert S} \approx &\frac{\pi_0^2}{\pi_0^2 + 2p\eta \pi_0} \ket{0} \bra{0} \\
		&+ \frac{2p \eta \pi_0}{\pi_0^2 + 2p\eta \pi_0} \ket{1} \bra{1} \\
		& + \frac{2p^2 \eta \pi_0}{\pi_0^2 + 2p\eta \pi_0} \ket{2} \bra{2}.
\end{split}
\end{equation*}
When the Stokes detection probability is larger than the probability of dark count, i.e. $\eta p \gg \pi_0$, this reduces
to the expression
\begin{equation}\label{app:rhobS}
	\op{\rho}_\t{b\vert S} \approx \frac{\pi_0}{2\eta p} \ket{0} \bra{0}
		+ \ket{1} \bra{1}
		+ p \ket{2} \bra{2}.
\end{equation}

\section{Time-evolved state of vibrational mode}

The time evolution of the vibrational state may be described by the master equation,
\begin{equation}\label{eq:master}
\begin{split}
	\dot{\op{\rho}}(t) = i\Omega_m [\rho(t), \op{b}^\dagger \op{b}] 
		&+ (\bar{n}+1)\Gamma_m\, \mathcal{D}[\op{b}] \op{\rho} \\ 
		&+ \bar{n}\Gamma_m\, \mathcal{D}[\op{b}^\dagger]\rho,
\end{split}
\end{equation}
where, $\Omega_m, \Gamma_m$ are the vibrational frequency and energy decay rate respectively, $\bar{n}$ is the average 
thermal occupation of a vibrational bath at frequency $\Omega_m$ and temperature $T$ given by the Bose distribution, and,
\begin{equation*}
	\mathcal{D}[\op{X}]\op{\rho} \equiv \op{X}\op{\rho} \op{X}^\dagger - \frac{1}{2}\{\op{X}^\dagger \op{X}, \op{\rho} \},
\end{equation*}
is the Lindblad super-operator modeling dissipation, with the jump operator $\op{X}$.

Since we are only interested in the properties of the Fock states of the vibrational mode, we consider 
the evolution of its number distribution $P_k(t)\equiv \bra{k}\op{\rho}(t)\ket{k}$. Equations of motion
for $P_k$ can be obtained by projecting \cref{eq:master} onto the Fock states, resulting in,
\begin{equation}\label{app:Pk}
\begin{split}
	\dot{P}_k(t) = \Gamma_m (\bar{n}+1) & \left[(k+1)P_{k+1}(t) - k P_k(t) \right] \\
		\quad+ \Gamma_m \bar{n} & \left[k P_{k-1}(t) -(k+1)P_k (t) \right].
\end{split}
\end{equation}
These represent a series of coupled rate equations for each component $P_k(t)$ of the phonon number distribution.

The rate equations can be solved exactly as follows. 
In order to decouple the various phonon number components we work with its generating function,
\begin{equation}\label{eq:Qgen}
	Q(z,t) \equiv \sum_{n=0}^\infty z^n P_n(t),
\end{equation}
which is a continuous function of $(z,t)$ and conveys the same information about the state as $P_n$; 
in fact, $P_n(t)=(1/n!)[\partial_z^n Q(z,t)]_{z=0}$.
Multiplying \cref{app:Pk} by $z^k$, summing over $k$, and performing the sum on the right-hand side, 
we obtain a partial differential equation for $Q(z,t)$:
\begin{equation}
	\frac{\partial Q}{\partial t} = \Gamma_m(1-z)\left[-\bar{n}+(1+\bar{n}(1-z))\frac{\partial}{\partial z} \right]Q.
\end{equation}
Given an initial state diagonal in the Fock basis (such as the state relevant to us), 
it is equivalently characterized by its moment generating function, $Q_0(z)\equiv Q(z,0)$. 
This data makes the above partial differential equation well-posed. 
This problem can be solved using the method of characteristics, giving,
\begin{equation}\label{app:Qzt}
	Q(z,t) = \frac{Q_0\left(1-\frac{(1-z)e^{-\Gamma_m t}}{1+\bar{n}(1-z)(1-e^{-\Gamma_m t})} \right)}
		{1+\bar{n}(1-z)(1-e^{-\Gamma_m t})}.
\end{equation}
In the following sections we use this solution to obtain the dynamics of the heralded vibrational state's phonon number distribution (\cref{app:Pnt}), and conditional intensity correlation function $g^{(2)}_\t{b\vert S}$ (\cref{app:g2b}).

\subsection{Number distribution}\label{app:Pnt}

We are interested in the number distribution of the phonon state between the write and read operations.
The initial state (\cref{app:rhobS}) is characterized by the moment generating function 
\begin{equation}\label{app:Qz0}
	Q_0(z) = P_0(0) + z P_1(0) + z^2 P_2(0),
\end{equation}
corresponding to the number distribution,
\begin{equation*}
\begin{split}
	P_0(0) &\approx \frac{\pi_0}{2\eta p} \lesssim 10^{-4}, \\
	P_1(0) &\approx 1, \\
	P_2(0) &\approx p \lesssim 10^{-2}.
\end{split}
\end{equation*}
The ambient thermal bath is well-approximated by a mean occupation 
$\bar{n} = 0$ (the vibrational mode oscillating at $\Omega_m \approx 2\pi \cdot 40\, \t{THz}$ at 
room temperature sees a bath of thermal occupation $\bar{n}\approx 2 \times 10^{-3}$). 
Under this approximation (\cref{app:Qzt}),
\begin{equation*}
	Q(z,t) = Q_0 (1-e^{-\Gamma_m t}(1-z)).
\end{equation*}
Inserting \cref{app:Qz0} into the right-hand side and collecting terms of equal powers in $z$, 
the required number distributions can be read-off as the corresponding factors; this gives,
\begin{equation}
\begin{split}
	P_0(t) &= P_0(0) + (1-e^{-\Gamma_m t})P_1(0) +(1-e^{-\Gamma_m t})^2 P_2(0) \\
	P_1(t) &= e^{-\Gamma_m t} P_1 (0) + 2e^{-\Gamma_m t}(1-e^{-\Gamma_m t})P_2(0) \\
	P_2(t) &= e^{-2\Gamma_m t} P_2 (0) .	
\end{split}
\end{equation}
These solutions are the ones plotted in Fig. 4c of the main text.

\subsection{Intensity correlation}\label{app:g2b}

In order to calculate the intensity correlation of the vibrational mode $g^{(2)}_\t{b\vert S}$, we refrain
from making the $\bar{n}=0$ approximation so as to recover the expected bunching behaviour of a thermal
state at long times. Further, we use the property that $Q(1-u,t)$, considered as a function of $u$, is
a generating function for the factorial moments of the number distribution \cite{mandel1995}. Specifically,
the two lowest order factorial moments are given by,
\begin{equation*}
\begin{split}
	\mean{\op{n}^{(1)}}_\t{\vert S} &\equiv \mean{\op{b}^\dagger \op{b}} _\t{\vert S}
		= \left[-\partial_u Q(1-u,t)\right]_{u=0} \\
	\mean{\op{n}^{(2)}} _\t{\vert S} &\equiv \mean{\op{b^\dagger} \op{b}(\op{b^\dagger}\op{b}-1)} _\t{\vert S}
		= \left[ (-\partial_u)^2 Q(1-u,t)\right]_{u=0},
\end{split}
\end{equation*}
which are related to the intensity correlation as, 
\begin{equation}\label{eq:g2def}
	g^{(2)}_\t{b\vert S} = \frac{\mean{\op{n}^{(2)}}_\t{\vert S}}{\mean{\op{n}^{(1)}}^2_\t{\vert S}}. 
\end{equation}
Using \cref{app:Qzt}, and using the approximation, $P_{1}(0)\approx 1 $, the intensity
correlation is given by,
\begin{equation}\label{app:g2bEq}
	g^{(2)}_\t{b\vert S} \approx \frac{2}{P_1(0)}\left[1-\frac{1}{(1+\bar{n}(e^{\Gamma_m t}-1))^2} \right].
\end{equation}
This equation, in conjunction with the phenomenological model for excess anti-Stokes noise (\cref{app:alpha})
is used in Fig. 4b of the main text.

\section{Model for noise in heralded intensity correlation}

The measured phonon intensity correlation at zero time delay, i.e. the  sub-Poisson character of the phonon Fock state
inferred via $\alpha (0) = g^{(2)}_\t{b\vert S}$, is limited by noise in the anti-Stokes field. 
We conjecture that this noise arises from a combination of the thermal phonon population and from spontaneous four-wave mixing generated by the intense read pulse \cite{kasperczyk2015,anderson2018}.
We refer all added noise to the phonon, and model its effect as the noiseless heralded phonon mode ($\op{b}$) interacting 
with a fictitious thermal phonon mode ($\op{b}_T$) via a beam-splitter type interaction,
\begin{equation}\label{eq:BSnoise}
	\op{b}' = \sqrt{1-\epsilon} \; \op{b} +\sqrt{\epsilon} \; \op{b}_T,
\end{equation}
to produce an effective noisy mode ($\op{b}'$).
The intensity correlation of this noisy mode is,
\begin{align}\label{eq:g2noise}
	g^{(2)}_{b'} &=\frac{(1-\epsilon)^2 n^2\, g^{(2)}_{b} + \epsilon^2 n_T^2\, g^{(2)}_{b_T} + 2\epsilon(1-\epsilon) n n_T}
		{\left[(1-\epsilon)n +\epsilon n_T \right]^2} \\
	&\approx g^{(2)}_{b} + 2\frac{\epsilon n_T}{(1-\epsilon)n};
\end{align}
here we have defined, $n \equiv \mean{\op{b}^\dag \op{b}}$, and, $n_T \equiv \mean{\op{b}^\dag_{T}\op{b}_{T}}$. 
The last approximation is an expansion to first order in the small parameter $\frac{\epsilon n_T}{(1-\epsilon)n}$, 
which models the noise-to-signal ratio. 
The effect of noise is thus an offset on the heralded intensity correlation parameter, i.e.,
\begin{equation}\label{app:alpha}
	\alpha (t) \approx g^{(2)}_{b} + \alpha_0,
\end{equation}
where, $\alpha_0 \equiv 2(S/N)^{-1}$, with $S/N$ the signal-to-noise ratio in the anti-Stokes detector. 

\subsection*{Independent estimate of the offset}

In the following, we provide an independent estimate of $\alpha_0$. 
Indeed, it can be shown that the normalized Stokes -- anti-Stokes intensity correlation is the anti-Stokes signal-to-noise ratio conditioned on Stokes detection. 
To see this, we introduce the following photon-counting probabilities:
\begin{itemize}
\item $p_N$, 
the probability of detecting a photon on one of the anti-Stokes detectors when the write pulse is off (or equivalently when the write-read delay is much longer than the phonon coherence time). This is the ``unconditioned'', or total, noise. 
\item $p_S$, 
the probability of Stokes detection, i.e. of successful heralding after a write pulse.
\item $p_C$,
the probability of a Stokes -- anti-Stokes coincidence, i.e. the probability of a heralded anti-Stokes detection event. 
\end{itemize} 

Note that $p_C$ times the repetition rate $\Gamma_R$ is the total heralded detection rate (in Hz) including signal and noise, i.e. $p_C\times \Gamma_R = S+N$. (Here we assume a fixed photon flux, allowing us to quantify
signal and noise in frequency units.)
The product $p_N \times p_S$ is directly accessible by measuring the probability of coincidence for very long (compared to 
the phonon lifetime) write-read delay; it is the probability of accidental coincidences.
This probability times the repetition rate is just the conditional noise, i.e. $p_N \times p_S \times \Gamma_R = N$.
Finally, the measured Stokes -- anti-Stokes intensity correlation is related to these quantities by the relation,
$$g^{(2)}_\t{S,AS}=\frac{p_C}{p_N p_S}=\frac{S+N}{N}\approx \frac{S}{N}=\frac{1}{\alpha_0/2}$$
where the last approximation is valid for $S/N \gg 1$.

Armed with this relation between $\alpha_0$ and $g_\t{S,AS}^{(2)}$, and the data for the latter 
$g_\t{S,AS}^{(2)} \approx 30$ (from Fig. 4a of main text), we estimate that, $\alpha_0 = 2 \times 0.033$, which is 
consistent with the value used ($\alpha_0 = 2\times 0.04$) to fit the data in Fig. 4b.

\section{Effect of loss in heralded intensity correlation}\label{app:detLoss}

\begin{figure}[t!]
	\centering
	\includegraphics[width=8cm]{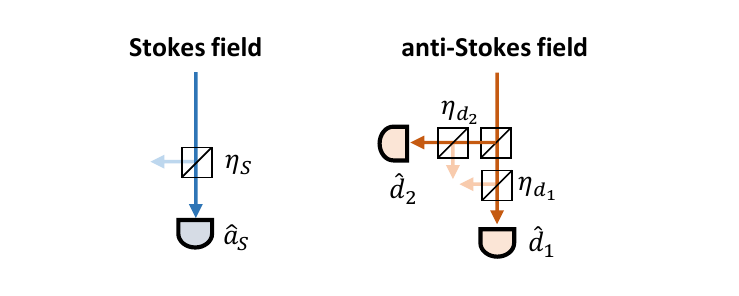}
	\caption{\label{fig:DetectionLosses} 
	Diagram showing the introduction of a beam-splitter of transmission $\eta_x$ before the detection of each mode.
}\end{figure}

The parameter $\alpha$ is not in general equivalent to $g^{(2)}_\t{AS|S}$ (or $g^{(2)}_\t{b|S}$); they are only equal in the limit of low detection efficiency in the Stokes detector. We illustrate this by considering the simplified example in \Cref{fig:DetectionLosses}, where beam splitters of transmission $\eta_S$, $\eta_{d_1}$, $\eta_{d_2}$ are added before the Stokes, $d_1$, and $d_2$ ideal detectors, respectively.

By assuming that $P(1,1) \gg P(2,2) \gg P(n>2,n>2)$ and using the beam-splitter input-output relations we can find the detection probability (per repetition of the experiment) for each detector by computing $P(n,n)$ for the different number states; this gives,
\begin{equation}\label{event_prob}
\begin{gathered}
N_{a_S} \simeq \eta_S P(1,1)
\\
N_{d_1,a_S} \simeq \frac{1}{2} (\eta_S) (\eta_{d_1}) P(1,1)
\\
N_{d_2,a_S} \simeq \frac{1}{2} (\eta_S) (\eta_{d_2}) P(1,1)
\\
N_{d_1,d_2,a_S} \simeq \frac{1}{2} (2\eta_S - \eta_S^2) (\eta_{d_1}) (\eta_{d_2}) P(2,2).
\end{gathered}
\end{equation}
Using the definition of $\alpha$ in the main text,
\begin{equation}\label{alpha_prob}
\alpha \simeq (4-2\eta_S) \frac{P(2,2)}{P(1,1)} = (4-2\eta_S)p
\end{equation}
It is clear that the detection losses due to $\eta_{d_{1,2}}$ do not affect the outcome of the measurement, and it is 
only the Stokes detection efficiency ($\eta_S$) which plays a role.
Specifically, for perfect Stokes detection ($\eta_S=1$), $\alpha = 2p = \frac{1}{2} g^{(2)}_\t{AS|S}$, while for low detection efficiency ($\eta_S \ll 1$), $\alpha \simeq 4p = g^{(2)}_\t{AS|S}$.

\section{Effect of pulse power on correlations}

The physics of our experiment is formally the same as that of spontaneous parametric down conversion (SPDC) in a nonlinear crystal with a $\chi^{(2)}$ susceptibility, once we have identified the Stokes (from the write pulse) and vibrational modes to the signal and idler modes in SPDC. Following the read pulse, we showed that the correlations between Stokes and vibrational modes are mapped onto those between Stokes and anti-Stokes modes (from the write and read pulse, respectively). The read operation can therefore be modeled as a direct detection of the phonon number in the vibrational mode, but with a reduced efficiency. 

It is therefore justified to use the model developed by Sekatski \textit{et al} \cite{sekatski2012} to understand our data taken at short write--read delay (Fig.~3), when we can neglect the loss of coherence in the vibrational mode. In particular we adapted the calculations leading to eqs.~(24) and (29) of Ref.~\cite{sekatski2012} to produce the blue lines in Fig.~3 of the main text, for the main panel and inset, respectively. More specifically, the heralded second order anti-Stokes correlation is expressed by
\begin{widetext}
\begin{equation}\label{cond_g2}
	\alpha(0) = \frac{1-2(1-\pi_\t{AS}) \zeta(1-\eta_\t{AS}/2) + 
		(1-\pi_\t{AS})^2 \zeta(1-\eta_\t{AS})}{(1-(1-\pi_\t{AS}) \zeta(1-\eta_\t{AS}/2))^2} 
\end{equation}
\end{widetext}
where 
\begin{equation}
\zeta(x)=\frac{1}{K} \left(\frac{1-p}{1-p x}-\frac{(1-\pi_0)(1-p)}{1-p(1-\eta_\t{S})x}\right).
\end{equation}
and
\begin{equation}
K = 1-\frac{(1-\pi_0)(1-p)}{1-p(1-\eta_\t{S})}
\end{equation}

The expression for the Stokes--anti-Stokes correlation (inset of Fig.~3) is

\begin{figure}[t!]
	\centering
	\includegraphics[width=8cm]{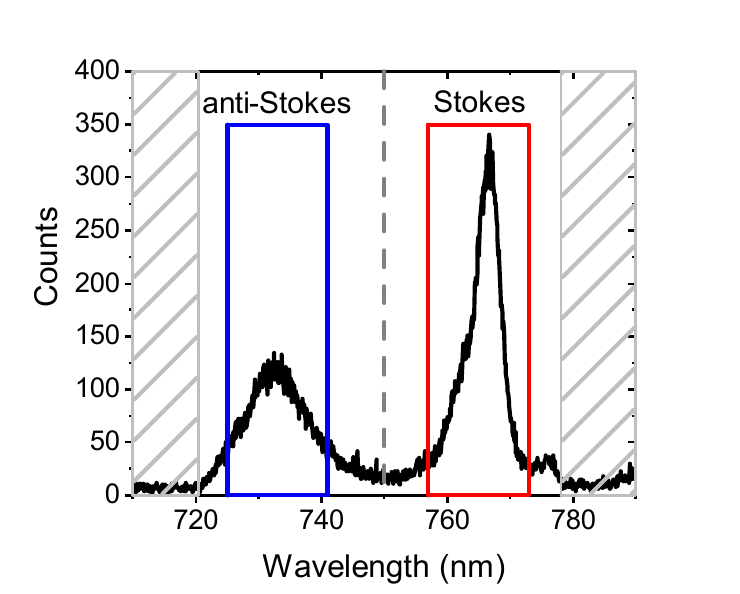}
	\caption{\label{fig:Spectrum} 
	{Spectrum of the Stokes and anti-Stokes signal in our experiments taken before the single mode fiber. The shaded regions at the edges represent the spectral regions cut by the long- and short-pass filters. The dashed line in the center represents the position of the long pass filter used to separate the two signals. The colored rectangles show the regions that are selected by the tunable band pass filters before the single photon counters. Note that unlike on this schematic representation the edges of the filters are not so sharp. However, their transmission decreases steeply from $>95$\% down to $<10^{-3}$ within less than 5~nm.}
}\end{figure}

\begin{align} \label{gab}
g_\t{S,AS}^{(2)} = \frac{N_\t{S,AS}}{N_0 N_\t{AS}} 
\end{align}
where,
\begin{equation}
\begin{split}
	N_\t{i} &= 1-(1-\pi_\t{i})\left(\frac{(1-p)}{1-p(1-\eta_\t{i})}\right), \qquad (i=\t{S,AS}) \\
	N_\t{S,AS} &= N_0 + N_\t{AS} - 1+ \\
		& \qquad (1-\pi_0)(1-\pi_\t{AS}) \left(\frac{(1- p)}{1- p(1-\eta_\t{S})(1-\eta_\t{AS})}\right)
\end{split}
\end{equation}

In contrast to Ref.~\cite{sekatski2012}, the detectors used for the anti-Stokes field (which indirectly probes the vibrational mode $b$) must be modeled with a reduced detection efficiency $\eta_\t{AS}$ and a higher probability of noise count $\pi_\t{AS}$, compared to the Stokes detector, for which $\eta_\t{S} = 10\%$ is the estimated overall detection efficiency of our setup and $\pi_0 = 7.5\pw{-7}$ the noise probability due to dark counts. 

The measured anti-Stokes count rate without write pulse is used to obtain $\pi_\t{AS} = 3.1 \pw{-5}$; it includes all sources of noise (in particular, the thermal contribution to the anti-Stokes signal and the four-wave mixing background). The effective phonon detection efficiency $\eta_\t{AS}$ is estimated from the corresponding Raman cross-section in the write pulse, extracted as the ratio of Stokes detection probability against the write pulse energy. However, to obtain agreement between the model and our data, we had to multiply this estimated readout efficiency by a factor $0.3$, which can be explained by a reduced Raman cross-section at longer wavelength for the read pulse, and an imperfect mode overlap between the read pulse and the heralded vibrational mode. We obtain the value $\eta_\t{AS}=0.019$ for the experimental conditions of Fig.~3.

\section{Spectral signal and detection window}

As described in the main text, we use a combination of tunable interference filters to select the spectral regions for the Stokes and anti-Stokes detection. The spectrum measured before the SM fiber (after blocking the write and read laser pulses with long-pass and a short-pass filter, respectively) is shown in Figure \ref{fig:Spectrum}, together with the detection windows selected with tunable long-pass and band-pass filters before performing the single photon detection.

\section*{Acknowledgements} 
Funding for this research was provided by the Swiss National Science Foundation, (project number PP00P2-170684). K.S. acknowledges support from the Swiss National Science Foundation (project number 200021-162357). V.S. is supported by a Swiss National Science Foundation fellowship (project number P2ELP2-178231). 
This research has received funding from the European Research Council (ERC) under the European Union’s Horizon 2020 research and innovation programme (Grant agreement No. 820196).

\bibliography{Heralded_aS_AB_biblio}

\end{document}